%% file: infodist.tex
\documentclass[twoside]{article}
\usepackage{qic}
\usepackage{eepic}
\usepackage{epic}

\newcommand{\prf}{{\textbf Proof.}}
\newcommand{\qed}{\mbox{\rule{1.6mm}{4.3mm}}}
\newcommand{\density}[2]{|{#1}\rangle\langle{#2}|}
\newcommand{\ket}[1]{|{#1}\rangle}
\newcommand{\bra}[1]{\langle{#1}|}
\newcommand{\braket}[2]{\langle {#1} | {#2} \rangle}

\newcommand{\nproj}[1]{\frac{|{#1}\rangle\langle{#1}|}{\langle{#1}|{#1}\rangle}}
\newcommand{\cc}{ C} 
\newcommand{\vph}{\varphi}

\begin{document}
\centerline{\bf
INFORMATION VS. DISTURBANCE IN DIMENSION D}
\vspace*{0.035truein}
\centerline{\footnotesize
P. Oscar Boykin
}
\vspace*{0.015truein}
\centerline{\footnotesize\it
Department of Electrical and Computer Engineering,
University of Florida}
\baselineskip=10pt
\centerline{\footnotesize\it Gainesville, Florida 32611,
USA
}
\vspace*{10pt}
\centerline{\footnotesize
Vwani P. Roychowdhury}
\vspace*{0.015truein}
\centerline{\footnotesize\it Department of Electrical Engineering,
University of California, Los Angeles}
\baselineskip=10pt
\centerline{\footnotesize\it Los Angeles, California 90024, USA}
\vspace*{0.225truein}

\vspace*{0.21truein}

\abstracts{
We show that for Eve to get information in one basis about
a state, she must
cause errors in \emph{all} bases that are mutually unbiased to
that basis.  Our result holds in any dimension.  We also show that this
result holds for \emph{all functions of messages} that are encrypted
with a key.
}{}{}

\vspace*{10pt}

\vspace*{3pt}

\vspace*{1pt}\textlineskip    

\section{Introduction}
Ideal quantum key distribution (QKD) {\em with qubits}\cite{BB84} is known to
be secure\cite{Mayers96,LC98,bbbmr,sp00,gl03}, and the security proofs are based on what are called information-vs.-disturbance results. The basic QKD protocol involves the following steps: Alice transmits one of four possible states randomly chosen from $|0\rangle_{X}, |1\rangle_{X}, |0\rangle_{Z},$ and $|1\rangle_{Z}$, i.e., the basis vectors in the $X$ and $Z$ bases.  The basic 
information-vs.-disturbance result states that if the eavesdropper, Eve, obtains information about which basis vector was sent in for example, the $X$ basis, then she must introduce disturbance in the $Z$ basis. By disturbance, it is meant that if Bob made measurements to distinguish between the two states sent in the $Z$ basis, then he will observe errors.  Thus Alice and Bob can test a random subset of a transmitted block of qubits in the $Z$ basis and estimate the information that Eve has about those in the $X$ basis. If the error rate is small enough in the tested qubits (hence, Eve's information about the qubits in the $X$ basis is small enough), then Alice and Bob can use classical error correcting and amplification schemes to distill an informationally secure key from the qubits sent in the $X$ basis. 

In this paper, we consider a general setup involving $D$ dimensional quantum states, instead of the 2-dimensional systems considered in the QKD literature. The basic setup is as follows:  Alice sends states chosen randomly from among the basis vectors of a particular basis of the $D$ dimensional Hilbert space. She intends these states to act as the information states, i.e., the $\log D$ bits per transmitted state will be used to distill a final key. The natural questions that arise are (i) which set of states should the  ``test" states come from,  and (ii) what is the corresponding information-vs.-disturbance result for a $D$-dimensional space.  

We first extend some basic distinguishability bounds
found for qubits\cite{fuchs99} to $D$-level systems. That is, if a source $S$ outputs one of $n$ $D$-dimensional quantum states randomly, then we derive bounds on the mutual information between $S$ and any measurement output $E$,  only in terms of the properties of the quantum states generated by $S$. In other words, we bound the mutual information between the random variable representing which state was generated by $S$ and the random variable representing the output from a generalized measurement of the states output by $S$. These results are powerful because they only depend on the
source and not on any measurement done. We next apply these bounds on
distinguishability to relate the amount of information eavesdroppers can obtain to the disturbance they cause in the quantum state. In particular, we prove a generalized information-vs.-disturbance result:  if Eve gets information about which basis vector (from the chosen basis in $D$ dimensions) was sent by  Alice, then she must introduce {\em disturbance in any basis that is mutually unbiased to the basis chosen by Alice}. 

In terms of previous work, our results generalize those in \cite{bbbmr,boykin02}. We would also like to note that  
 QKD in dimension $3$ was studied in \cite{bm02,btb04}.  Security bounds for
individual cloning attacks in dimension $D$ have been reported\cite{ags03}.
More recently, qubit QKD techniques\cite{LC98,sp00} have been
generalized to prime dimensions\cite{Chau04}.
By contrast, our bounds {\em apply to any attack in any dimension}.
Also, this work further illuminates the relationship of mutually unbiased bases (MUBs)\cite{ivanovic81} to quantum cryptography.
 Previously, it was shown that the eigenvectors of
maximally commuting
quantum encryption operators form MUBs\cite{bbr02}.  Here we show
that when Eve tries to get information in one basis, she disturbs
\emph{all} MUBs.
Our result may be viewed as form of an uncertainty principle:
the more Eve knows about one basis, the more she disturbs
\emph{all} conjugate bases.

In addition to applying the above bounds and techniques to the security
of quantum keys, we also consider \emph{functions of messages
encrypted with those keys}.  If Alice and
Bob share a key $k$, it may be that Eve learns only exponentially little
information about $k$, but she may be able to learn a lot about
some function of a message $f(M)$, given the encrypted version of
that message $m+k$. In particular, consider the following setup:
Alice sends a random basis vector $\left| k\right\rangle$ belonging to a chosen basis to Bob. Alice next publicly announces she sent basis vector $\left| k\oplus m \right\rangle$, where $\oplus$ is the bitwise exclusive or (XOR) operation. Bob could then recover the encrypted message $m$. Now, we know that information of Eve about $k$ is bounded by the error she causes in any basis that is mutually unbiased to the chosen basis. How about a function $f(M)$ of the message? For example, Eve might be interested in only learning whether $m=0$ or not.  In a previous work\cite{boykin02}, it was shown that given
the encrypted message, $m + k$, the information
that Eve gets about any function of an encrypted n-bit message
$f(m)$, is bounded by the square root of the error Eve's
attack causes in the Hadamard transformed basis.
More recently, alternative and more general
solutions to this problem have been given
\cite{bhl04,rk04}.
In this work we extend
previous results\cite{boykin02}
beyond qubits to $d$-dimensional systems.  Also, we show
that Eve's information is bounded by the error she causes is \emph{any}
MUB.

This paper is structured as follows: Section \ref{sec:info_bounds}
gives various new bounds on distinguishability and classical information
accessible from quantum states; Section \ref{sec:qkd} applies these results
to obtain ``information-vs-disturbance" results for QKD; finally
in Section \ref{sec:sfunc} we show these results also hold for \emph{functions
of encrypted messages}
and not just for the keys themselves.

\section{Bound On Information For Any Source}
\label{sec:info_bounds}
In \cite{fuchs99}, many bounds are given on the distinguishability of two quantum
states.  In this section we generalize some of those to the distinguishability of $n$
quantum states.  Our setting is the following: A source outputs one of $n$ quantum
states.  The random variable representing the source is $S$ i.e., it is the identifier of the particular quantum state made available at the output and can be generated by purely classical means, such as flipping coins or spinning wheels.  A general measurement
is made on the state, which results in one of several measurement outcomes represented by
the random variable $E$.  We consider bounds on the mutual information $I(S;E)$ valid for any measurement, which
is to say, the bound will only be a function of the quantum states emitted by the source.   

The bounds here address the same problem as the well known Holevo bound\cite{Kholevo73},
which is:
\begin{equation}
\label{eq:holevo}
I(S;E) \le H(\rho) - \sum_s p_s H(\rho_s)
\end{equation}
where $H(\rho)$ is the Von-Neumann entropy of the density matrix $\rho$.
The main difference between the results of this section
and the Holevo bound is that these results deal
explicitly with a distance metric, namely the trace norm distance, between two density
matrices.  Using a simple distance metric allows a certain ease in proving the results in
Section \ref{sec:qkd}\footnote{We do believe, however, that it is possible to obtain similar results
by applying the purification techniques of Section \ref{sec:qkd} directly to the Holevo bound.}

In the appendix, we review certain previously published \cite{fuchs99,boykin02} bounds
on distinguishability of quantum states.
As we will see later in the paper, this allows us to derive
the fundamental information vs. disturbance results that are at work in quantum security protocols.
Additionally, these results give an important insight into the robustness of
the trace norm as a metric bound for information.

We begin by developing a lower bound on entropy and then applying that
bound to the mutual information.
\begin{lemma}
\label{lemm:gen-entropy-bound}
For any random variable $X'$ with each probability ${p_i}' \le 1/2$:
\begin{eqnarray*}
H(X)\ge H(X') - \sum_i \log(\frac{1}{{p_i}'})|p_i - {p_i}'| 
\end{eqnarray*}
\end{lemma}
\prf
$H(X)=-\sum_i p_i \log p_i$, so if we define $f(p_i)\equiv -p_i \log p_i$,
we see that $H(X) = \sum_i f(p_i)$.  See that $f$ is concave
and is zero at $p_i=0,1$; thus lemma \ref{lemm:entropy-bound} applies:
\begin{eqnarray*}
f(p_i)&\ge&f({p_i}') - \frac{f({p_i}')}{{p_i}'}|p_i - {p_i}'|
\end{eqnarray*}
Plugging this into the definition of entropy:
\begin{eqnarray*}
H(X)&=&\sum_i f(p_i)\\
&\ge& \sum_i (f({p_i}') - \frac{f({p_i}')}{{p_i}'}|p_i - {p_i}'|)\\
&=& H(X') - \sum_i \log(\frac{1}{{p_i}'})|p_i - {p_i}'|
\end{eqnarray*}
\qed

\begin{lemma}
\label{lemm:nbit_mi}
For any source S that outputs $s$ with probability $p_s$ such that
$p_s \le 1/2$, the mutual information is bounded:
\begin{eqnarray*}
I(S;E)&\le&\sum_s p_s \log(\frac{1}{p_s})\sum_e |p(e|s) - p(e)|
\end{eqnarray*}
\end{lemma}
\prf
Make use of lemma \ref{lemm:gen-entropy-bound}:
\begin{eqnarray*}
I(S;E)&=&H(S) - H(S|E)\\
&=&H(S) - \sum_e p_e H(S|E=e)\\
&\le&H(S) - \sum_e p_e \left( H(S) -
\sum_s \log(\frac{1}{p_s})|p(s|e) - p_s|\right)\\
&=& \sum_e p_e \sum_s \log(\frac{1}{p_s})|p(s|e) - p_s|\\
&=& \sum_e \sum_s p_s\log(\frac{1}{p_s})|\frac{p(e)p(s|e)}{p_s} - p(e)|\\
&=& \sum_e \sum_s p_s\log(\frac{1}{p_s})|p(e|s) - p(e)|.
\end{eqnarray*}
\qed
\begin{lemma}
\label{lemm:nbitSD}
If a source $S$ outputs quantum states $\rho_i$ with
probabilities $p_i$ with $p_i \le 1/2$, then mutual information
between this source and the output of any measuring device $E$ is bounded:
\begin{eqnarray*}
I(S;E) &\le& \sum_s p_s \log(\frac{1}{p_s})Tr|\rho_s - \sum_s p_s \rho_s|.
\end{eqnarray*}
\end{lemma}
\prf
Define the notation $\rho = \sum_s p_s \rho_s$.
Starting from lemma \ref{lemm:nbit_mi},
we use the definition of a POVM to replace $p(e|s)$ with
$Tr(E_e \rho_s)$:
\begin{eqnarray*}
I(S;E) &\le& \sum_e \sum_s p_s\log(\frac{1}{p_s})|p(e|s) - p(e)|\\
&=& \sum_e \sum_s p_s\log(\frac{1}{p_s})|Tr(E_e \rho_s) - Tr(E_e \rho)|\\
&=& \sum_e \sum_s p_s\log(\frac{1}{p_s})|Tr(E_e(\rho_s - \rho))|
\end{eqnarray*}
Using the same facts about POVMs as in lemma \ref{lemm:1bitSD}, one can show that
\begin{eqnarray*}
\sum_e |Tr(E_e(\rho_s - \rho))| &\le& Tr|\rho_s - \rho|.
\end{eqnarray*}
Hence, we have:
\begin{eqnarray*}
I(S;E) &\le& \sum_s p_s\log(\frac{1}{p_s})Tr|\rho_s - \rho|.
\end{eqnarray*}
\qed
\begin{corollary}
\label{co:sd_equal_p}
If a source $S$ outputs one of $n$ quantum states $\rho_i$ with
probability $1/n$, then mutual information
between this source and the output of any measuring device $E$ is bounded:
$I(S;E)\le \log n \sum_s \frac{1}{n}|\rho_s - \rho|$.
\end{corollary}
\prf
For all $n\ge 2$, then $1/n \le 1/2$, hence lemma \ref{lemm:nbitSD} applies:
\begin{eqnarray*}
I(S;E) &\le& \sum_s p_s\log(\frac{1}{p_s})Tr|\rho_s - \rho|\\
&=& \log n \sum_s \frac{1}{n}Tr|\rho_s - \rho|
\end{eqnarray*}
\qed

Now we have a basic lemma in hand which gives an upper bound on the information
any measurement device can get from any source, purely in terms of the quantum states
emitted from that source.  In the next section, we will model the eavesdropping process
as a source of quantum states for Eve.  Eve is free to measure states in any way,
but using the previous lemma, we have an upper bound on how much information she may
obtain.

\section{Security of Quantum Key Distribution}
\label{sec:qkd}
We now have the tools necessary in order to derive an {\em information
theoretic counterpart to the Heisenberg uncertainty principle}.  This result
is the basis for quantum security results in \cite{bbbmr}.
Quantum key distribution (QKD) is directly related to the setup we considered
in the previous section. In general, in a QKD setup Alice has the source $S$ that outputs one of $n$ quantum states; Alice transmits the output state over a quantum channel to Bob. This quantum channel, however, can belong to the eavesdropper Eve, who can perform any operation that quantum mechanics allows. Figure \ref{fig:basic} gives a schematic of the most general attack
that Eve might perform.  From her perspective, she has access to a source, and she can make any
measurement to get information about what was sent.  Bob thus receives a state that Eve has already processed and makes his own measurements using a fixed protocol that is known to everyone. Alice and Bob complete a block transmission of several output states of the source $S$, and then use classical communication over an open channel to distill a secret key. Eve can listen in as well on the classical channel, but cannot perform a person-in-the-middle attack on the classical channel, which will make the whole protocol trivially unsecured.  Such a classical channel can be easily implemented by message authentication, e.g., via previously shared secret bits between Alice and Bob.  

Security of the QKD schemes depend on the amount of mutual information between Alice's source, $S$, and Eve's measurement $E$ (i.e., $I(S;E)$ as considered in the previous section) when measured as a function of the disturbance that she causes to the state received by Bob. The intuition from 
quantum mechanics is that measurements will disturb the system; hence, Alice and Bob can use a random subset of the transmitted quantum states for testing purposes, and detect the error rate on this subset, and thereby infer how strongly has Eve attacked the whole block. The underlying result and assumption here is that if the error she causes is less than a threshold then so is the mutual information $I(S;E)$. They proceed with key distillation only if the test errors are below a pre-specified threshold. Next, one can use classical privacy amplification schemes to  show that as long as $I(S;E)$ is small enough (as implied by the disturbance), then one can make the mutual information between $E$ and a final distilled key as low as possible. These classical techniques involve the use of error correcting codes.

\begin{figure}
\begin{center}
\input{basic_attack.eepic}
\hbox{Fig.~\ref{fig:basic} Most general attack by an eavesdropper.}
\end{center}
\label{fig:basic}
\end{figure}

Thus, the derivation of an appropriate ``information vs. disturbance" result lies at the heart of all security proofs for QKD.
While it is clear what we mean by ``information," (as defined by the quantity $I(S;E)$), we have not yet quantified and defined what we mean by ``disturbance." In various security proofs of QKD, researchers have  adopted the following strategy: (i) In the protocol, the source $S$ outputs states chosen from the basis vectors belonging to two different bases, e.g., the $X$ and $Z$ bases. (ii) The information vs. disturbance results then refer to the information about which basis vector from one of the bases (e.g., $X$) was sent, and the disturbance caused in the second basis (e.g., $Z$).  That is, Eve cannot simultaneously get significant information about which basis vector was sent in one basis, without causing errors in Bob's inference about which basis vector was sent in the other basis. Thus for testing purposes, one could use the states in one of the bases and the observed error rate will put a bound on the information that Eve has about which basis vectors were sent in the other bases.

Specifically, Lo and Chau\cite{LC98} use an EPR based scheme and show 
(using the Holevo bound, equation \ref{eq:holevo}) that if the fidelity
between Alice and Bob is greater than $1-\delta$ for $R$ singlets, then Eve's
information about the final key is bounded by:
\begin{eqnarray*}
I &\le& -(1-\delta)\log(1-\delta) - \delta \log \frac{\delta}{2^{2R}-1}
\end{eqnarray*}
The above information-vs-disturbance result is used directly by Shor and
Preskill in their quantum code based proof\cite{sp00}.  Rather that deal with
the fidelity of singlets, Biham et. al.\cite{bbbmr} use trace-norm techniques to show
that Eve's information on each bit is bounded by the square root of the
probability that she would cause more than $\hat{v}/2$ errors had
Alice sent the bits in the opposite basis (X replaced with Z and
vice-versa), where $\hat{v}$ is the minimum distance between the privacy
amplification code and the error correction code.  The security of QKD
directly depends on the above results: Eve's information is
always bounded once Alice and Bob verify that their states have not been
greatly disturbed.

In this section, we generalize such information vs. disturbance bounds for states in any dimension $D$, and {\em also provide a natural choice of the bases} to be used in these results. At this point it is useful to define the concept of Mutually Unbiased Bases:

\noindent
{\bf Definition.} Let
$B_1=\left\{\ket{\varphi_1},\ldots,\ket{\varphi_D}\right\}$ and
$B_2=\left\{\ket{\psi_1},\ldots,\ket{\psi_D}\right\}$be two orthonormal
bases in the $D$ dimensional state space. They are said to be
{\bf mutually unbiased bases (MUB)}   
if  and only if
$\left\vert\braket{\vph_i}{\psi_j}\right\vert=\frac{1}{\sqrt{D}}$, for every
$i,j=1,\ldots,d$. A set $\left\{ {\cal B}_1,\ldots,{\cal B}_m\right\}$ of
orthonormal bases in $\cc^D$ is called a {\em set of mutually unbiased bases}
(a set of MUB) if each pair of bases ${\cal B}_i$ and ${\cal B}_j$ are
mutually unbiased.

Thus, given two MUB $B_1$ and $B_2$, we get $B_1B_2^{\dag}= H$, where $|H_{i,j}| =1/\sqrt{D}$, and $H$ is  a unitary matrix.  Hence, {\em $H$ can be regarded as a generalized Hadamard matrix} in dimension $D$, and the two bases are related by the transformation $B_1=HB_2$. We next derive a general theorem which shows that whatever the dimension,
if Eve gets information in one basis, she disturbs \emph{all} bases
which are MUBs of that basis. Since two MUB are related by a generalized Hadamard transformation, the result in Theorem 1 implies that retrieving information in one basis  causes disturbances in {\em all the conjugate bases}. 

Finally, it should be emphasized that we only consider a single $D$-dimensional
state.  This is not a limitation: any product of quantum states can be
thought of as a state in a larger dimensional space.  Thus, if we consider
standard BB84, $n$ 2-dimensional systems (bits) are sent.  In our
approach we would
consider that as one $2^n$ dimensional system.  The same applies for any
product of quantum states.
These results generalize those presented in \cite{boykin02}, which proved
the following theorem only for dimension $2^n$ and for one pair of bases
(the standard $Z$ and $X$ bases).
\begin{theorem}
\label{thm:ivd}
If Alice sends a randomly selected element from a $D$-dimensional basis (represented
by the random variable $A$)
to Bob, the
information Eve's measurement (represented by $E$) has about Alice's state is
bounded by the square root of
the probability that Eve would have caused errors in any MUB with respect to Alice's basis:
\begin{eqnarray*}
I(A;E)&\le& 4\log D\sqrt{P_{\widetilde{e}}}\ .
\end{eqnarray*}
\end{theorem}
\prf
We will use lemmas \ref{lemm:trace_out_bound} and
\ref{lemm:trace_bound_mixed_pure} and corollary \ref{co:sd_equal_p}.
Starting from corollary \ref{co:sd_equal_p} we see that:
$I(A;E)\le \log D \sum_i \frac{1}{D}|\rho_i - \rho|$.  Our approach will be to
bound this by introducing a purification\footnote{see definition
\ref{def:purification}} for $\rho_i$ (the state that Eve
holds when Alice sends $i$).  Using the purification and
lemma \ref{lemm:trace_out_bound} we can bound the original trace
norm distance. 

To attack the state sent to Bob, Eve attaches a probe in a fixed state (say
the $\ket{0}$ state) and applies a unitary operator.  She then passes Bob his
part, and does some generalized measurement on what she still holds.  We can
characterize this formally:
\begin{eqnarray*}
\ket{0}_{E}\ket{i}_A\stackrel{U}{\rightarrow}\sum_j \ket{E_{i,j}}\ket{j}
\end{eqnarray*}
We represent the MUB as:
\begin{eqnarray*}
\ket{\widetilde{i}}&\equiv& \sum_j H_{ji}\ket{j}
\end{eqnarray*}
With $H$ being a generalized Hadamard matrix on these $D$-dimensional basis:
$|H_{ji}|=\frac{1}{\sqrt{D}}$.
Applying this to Eve's attack, we obtain:
\begin{eqnarray*}
\ket{0}_{E}\ket{\widetilde{i}}_A\stackrel{U}{\rightarrow}\sum_j
\ket{\widetilde{E_{i,j}}}\ket{\widetilde{j}}
\end{eqnarray*}
where $\ket{\widetilde{E_{i,j}}}\equiv \sum_{i',j'}H_{i' i}H^{*}_{j' j}\ket{E_{i',j'}}$.

From the axioms of quantum mechanics, we know that
if Alice sends $\ket{i}$ the probability that Bob will measure $\ket{j}$ is
$P(j|i)=\braket{E_{i,j}}{E_{i,j}}$.  Similarly, if Alice sends $\ket{\widetilde{i}}$ Bob
will measure $\ket{\widetilde{j}}$ with probability
$\widetilde{P}(j|i)=\braket{\widetilde{E_{i,j}}}{\widetilde{E_{i,j}}}$.

We are now prepared to compute the probability that there are no errors
in the MUB:
\begin{eqnarray}
P_0&\equiv&\sum_i p(i)\widetilde{P}(i|i)\nonumber\\
&=&\frac{1}{D}\sum_i \braket{\widetilde{E_{i,i}}}{\widetilde{E_{i,i}}}\nonumber\\
&=&\frac{1}{D}\sum_i\sum_{k,l,k',l'}H^{*}_{l i} H_{k i} H_{l' i} H^{*}_{k' i}
    \braket{E_{l,k}}{E_{l',k'}}\nonumber\\
&=&\frac{1}{D}\sum_{k,l,k',l'}\braket{E_{l,k}}{E_{l',k'}}
    \sum_i H^{*}_{l i} H_{k i} H_{l' i} H^{*}_{k' i}\label{eq:p0_sum}
\end{eqnarray}

When Eve's states are considered without Bob, her state will look like $\rho_i
= \sum_j \density{E_{i,j}}{E_{i,j}}$.
Now we will define a purification for Eve's states that will allow us to compute
a bound on $P_0$. 
We assume that Eve holds 
\begin{equation}
\label{eq:phi_i}
\ket{\phi_i}\equiv\sum_j\ket{E_{i,j}}_1\ket{\psi^i_j}_2\nonumber
\end{equation}
where $\ket{\psi^i_j}$ is an orthonormal basis for each choice of $i$.
Due to the orthonormality of $\ket{\psi^i_j}$,
$\ket{\phi_i}$ is a purification of $\rho_i$ because
$Tr_2\density{\phi_i}{\phi_i} = \rho_i$.
We also define the
generalized Hadamard transform of these states:
\begin{equation}
\ket{\widetilde{\phi_j}}\equiv\sum_i H^{*}_{i j}\ket{\phi_i}
\ .
\end{equation}
The Hadamard 
transform is unitary, so see that
$\ket{\phi_i} = \sum_j H_{i j}\ket{\widetilde{\phi_j}}$.
It should be noted that our purification
$\ket{\phi_i}$ for Eve's states is not orthonormal or normalized.  In fact,
this is a property of which we will make use in order to get a bound.  
We now calculate
the norm of the $\ket{\widetilde{\phi_0}}$ and see that
with the proper choice of $\ket{\psi^i_j}$ that it is proportional to the
probability that there
was no error, $P_0$:
\begin{eqnarray}
\braket{\widetilde{\phi_0}}{\widetilde{\phi_0}}&=&
\sum_{l,l'}H_{l 0} H^{*}_{l' 0}\braket{\phi_l}{\phi_{l'}}\nonumber\\
&=&\sum_{l,l'}\sum_{k,k'}H_{l 0} H^{*}_{l' 0}
\braket{E_{l,k}}{E_{l',k'}}\braket{\psi^{l}_k}{\psi^{l'}_{k'}}
\label{eq:phi_bar_0}
\end{eqnarray}
At this point we will parameterize $\ket{\psi^l_k}$:
\begin{eqnarray*}
\ket{\psi^l_k} &=& \sum_i \alpha_{l k i} \ket{i}
\end{eqnarray*}
with any choice of $\alpha_{l k i}$ so long as
$\braket{\psi^l_{k'}}{\psi^l_k}=\delta_{k' k}$.
In order to match equation \ref{eq:p0_sum} with
equation \ref{eq:phi_bar_0}, we choose
\begin{equation}
\alpha_{l k i} = \frac{H_{l i} H^{*}_{k i}}{H^{*}_{l 0}} \ .
\end{equation}
To see that our choice of $\alpha_{l k i}$ is valid,
recall that $|H_{i j}|^2 = 1/D$ and simply compute
\begin{eqnarray*}
\braket{\psi^l_{k'}}{\psi^l_{k}}
&=& \sum_i \alpha^{*}_{l k' i} \alpha_{l k i}\\
&=&\frac{1}{|H_{l 0}|^2}
\sum_i |H_{l i}|^2 H_{k' i} H^{*}_{k i}\\
&=& \sum_i H_{k' i} H^{*}_{k i}\\
&=& \delta_{k' k}
\end{eqnarray*}
which is what we need to show to make equation \ref{eq:phi_i} a valid
purification.
With the above choice, equation \ref{eq:phi_bar_0} becomes
\begin{eqnarray*}
\braket{\widetilde{\phi_0}}{\widetilde{\phi_0}}&=&\sum_{l,l'}\sum_{k,k'}H^{*}_{l 0} H_{l' 0}
\braket{E_{l,k}}{E_{l',k'}}\braket{\psi^{l}_k}{\psi^{l'}_{k'}}\\
&=&\sum_{k,l,k',l'}\braket{E_{l,k}}{E_{l',k'}}
    \sum_i H^{*}_{l i} H_{k i} H_{l' i} H^{*}_{k' i}\\
&=& D P_0 \ .
\end{eqnarray*}
Thus we have related the norm of $\ket{\widetilde{\phi_0}}$ to the probability
that there are no errors
\footnote{If the Hadamard transform is isomorphic to a group such that
$H_{i k}H_{j k} = H_{i+j,k}\frac{1}{\sqrt{D}}$ and
$H_{i k}H^{*}_{j k} = H_{i-j,k}\frac{1}{\sqrt{D}}$
we can show that the probability of an error $e$ in the Hadamard transformed
basis (i.e. Alice sends $i$ but Bob receives $i+e$ averaged over all $i$),
is $P_e = \braket{\widetilde{\phi_e}}{\widetilde{\phi_e}}/D$.
In this case, $\ket{\psi^i_j}=\ket{\widetilde{i-j}}$.  Indeed, this is the case
for the standard Sylvester type Hadamard matrices.
}
in the MUB.

Define ${\rho_i}'\equiv
\density{\phi_i}{\phi_i}$ and ${\rho}'\equiv \frac{1}{D}\sum_i\rho_i$.
Now we compute $\bra{\widetilde{\phi_0}}{\rho}'\ket{\widetilde{\phi_0}}$:
\begin{eqnarray*}
\bra{\widetilde{\phi_0}}{\rho}'\ket{\widetilde{\phi_0}}
&=&\sum_i \frac{1}{D}|\braket{\widetilde{\phi_0}}{\phi_i}|^2\\
&=&\sum_i
\frac{1}{D}|\bra{\widetilde{\phi_0}}\sum_j H_{i j}\ket{\widetilde{\phi_j}}|^2
\end{eqnarray*}
Since $|H^{*}_{i k}|^2 D = 1$, we can rewrite the above as:
\begin{eqnarray*}
\bra{\widetilde{\phi_0}}{\rho}'\ket{\widetilde{\phi_0}}
&=& D \sum_i \frac{1}{D}
|H^{*}_{i k}\bra{\widetilde{\phi_0}}\sum_j H_{i j}\ket{\widetilde{\phi_j}}|^2
\end{eqnarray*}
Since $f(x)=|x|^2$ is convex, then $|\sum_i p_i x_i|^2 \le \sum_i p_i |x_i|^2$.
\begin{eqnarray*}
\bra{\widetilde{\phi_0}}{\rho}'\ket{\widetilde{\phi_0}}
&=& D \sum_i \frac{1}{D}
|H^{*}_{i k}\bra{\widetilde{\phi_0}}\sum_j H_{i j}\ket{\widetilde{\phi_j}}|^2\\
&\ge& D | \sum_i \frac{1}{D}
H^{*}_{i k}\bra{\widetilde{\phi_0}}\sum_j H_{i j}\ket{\widetilde{\phi_j}} |^2\\
&=& D | \frac{1}{D}
\bra{\widetilde{\phi_0}}\sum_j\sum_i H^{*}_{i k} H_{i j}\ket{\widetilde{\phi_j}} |^2\\
&=& D | \frac{1}{D}
\bra{\widetilde{\phi_0}}\sum_j\delta_{k j}\ket{\widetilde{\phi_j}} |^2\\
&=& D | \frac{1}{D}
\braket{\widetilde{\phi_0}}{\widetilde{\phi_k}} |^2\\
&=& \frac{1}{D} |
\braket{\widetilde{\phi_0}}{\widetilde{\phi_k}} |^2
\end{eqnarray*}
We can set $k$ to any value we like, in particular $k=0$.  We have previously
shown that $\braket{\widetilde{\phi_0}}{\widetilde{\phi_0}} = D P_0$, putting this
together:
\begin{eqnarray*}
\bra{\widetilde{\phi_0}}{\rho}'\ket{\widetilde{\phi_0}} &\ge&
\frac{1}{D}|\braket{\widetilde{\phi_0}}{\widetilde{\phi_0}}|^2\\
&=& \braket{\widetilde{\phi_0}}{\widetilde{\phi_0}} P_0\\
\frac{ \bra{\widetilde{\phi_0}}{\rho}'\ket{\widetilde{\phi_0}}  }
{ \braket{\widetilde{\phi_0}}{\widetilde{\phi_0}} } &\ge& P_0
\end{eqnarray*}

We are now ready to prove the theorem.
Since $Tr_2(\rho_i')=\rho_i$ and $Tr_2(\rho')=\rho$ we may
apply lemma \ref{lemm:trace_out_bound}.
We will see that we may introduce an intermediate pure state to
make the bounding of the information easier.  The pure state we
will use is $\nproj{\widetilde{\phi_0}}$. 
Starting with corollary \ref{co:sd_equal_p}:
\begin{eqnarray*}
I(A;E)&\le& \log D \sum_i \frac{1}{D}|\rho_i - \rho|\\
&\le& \log D \sum_i \frac{1}{D}|{\rho_i}' - {\rho}'|\\
&=& \log D \sum_i \frac{1}{D}|{\rho_i}' 
- \nproj{\widetilde{\phi_0}} + \nproj{\widetilde{\phi_0}} - {\rho}'|\\
&\le& \log D \sum_i \frac{1}{D}(|{\rho_i}' -
\nproj{\widetilde{\phi_0}}| + |\nproj{\widetilde{\phi_0}}
- {\rho}'|)\\
&\le& \log D \sum_i \frac{1}{D} \left(2\sqrt{1 -
\frac{\bra{\widetilde{\phi_0}}{\rho_i}'\ket{\widetilde{\phi_0}}}
{\braket{\widetilde{\phi_0}}{\widetilde{\phi_0}}} } + 2\sqrt{1 -
\frac{\bra{\widetilde{\phi_0}}{\rho}'\ket{\widetilde{\phi_0}}}
{\braket{\widetilde{\phi_0}}{\widetilde{\phi_0}}} } \right)\\
&=& 2\log D \left(\sqrt{1 -
\frac{\bra{\widetilde{\phi_0}}{\rho}'\ket{\widetilde{\phi_0}}}
{\braket{\widetilde{\phi_0}}{\widetilde{\phi_0}}} } + \sum_i \frac{1}{D}
\sqrt{1 - \frac{\bra{\widetilde{\phi_0}}{\rho_i}'\ket{\widetilde{\phi_0}}}
{\braket{\widetilde{\phi_0}}{\widetilde{\phi_0}} } }\right)\\
&\le& 2\log D \left(\sqrt{1 -
\frac{ \bra{\widetilde{\phi_0}}{\rho}'\ket{\widetilde{\phi_0}} }
{ \braket{\widetilde{\phi_0}}{\widetilde{\phi_0}} }
} +
\sqrt{1 - 
\frac{ \bra{\widetilde{\phi_0}}(\sum_i \frac{1}{D}{\rho_i}')\ket{\widetilde{\phi_0}} }
{ \braket{\widetilde{\phi_0}}{\widetilde{\phi_0}} }
}\right)\\
&=& 4\log D \sqrt{1 -
\frac{\bra{\widetilde{\phi_0}}{\rho}'\ket{\widetilde{\phi_0}}}
{ \braket{\widetilde{\phi_0}}{\widetilde{\phi_0}} } }\\
&\le & 4\log D\sqrt{ 1 - P_0}
\end{eqnarray*}

Where $1-P_0 = P_{ \widetilde{ e } }$ is the probability that there is an error in the MUB, which proves the theorem.
\qed

The previous theorem is what gives security to quantum key distribution
schemes; however, we have only shown that QKD schemes are secure
if the errors caused in any MUB are extremely small.
Using quantum coding based approaches\cite{sp00}, we believe it is possible
to use the above theorem
to get a simple unconditional security proof that applies in dimension $D$.

In the following section, we will apply these same techniques to
show that Eve also cannot learn functions of messages.

\section{Security of Functions of Messages}
\label{sec:sfunc}
According to theorem \ref{thm:ivd},
if the fidelity Bob would have had in any
MUB is exponentially close to unity, then Eve's
information is exponentially low about which of the basis vectors in the chosen basis was sent. We will refer to the identifier of the basis vector sent by Alice as the {\em key}, and {\em Alice can use the key to encrypt a classical message}.   For example, after sending a basis vector $\left| k\right\rangle$ to Bob, Alice could publicly announce she sent basis vector $\left| k\oplus m \right\rangle$, where $\oplus$ is the bitwise exclusive or (XOR) operation. Bob could then recover the encrypted message $m$. 

The above mentioned information vs. disturbance result  does not address the question of what information
Eve might get about a \emph{function} of a message encrypted with that key.
Suppose Eve only wants to
know if the message has a particular value, i.e., she wants to learn the indicator function: $f(m)=1$ if $m=m_1$, else $f(m)=0$.  This function only has exponentially
little information about the message itself. To see this, 
suppose each of $d$ messages are equally likely, then
\begin{eqnarray*}
H(M)&=&\log d\\
H(f(M))&=&\frac{1}{d}\log d - (1-\frac{1}{d})\log(1-\frac{1}{d})\\
H(f(M)|M)&=& 0\\
I(f(M);M) &=& H(f(M)) \ .
\end{eqnarray*}
If $d$ is large, then $H(f(M))\approx \frac{1}{d}\log d$, but, $d = 2^{H(M)}$,
so $H(f(M)) \approx 2^{-H(M)} H(M)$.  Hence, in this case, Eve only
has to learn exponentially little information.  Since QKD
security proofs\cite{Mayers96,LC98,bbbmr,sp00,gl03} only
give exponentially strong security, it is not clear a priori that QKD will
be sufficient to prevent Eve from learning any function of the message.

The next
theorem will show that Eve must cause errors to {\em learn any function
of the message}, even if it has exponentially little information with
the message itself\footnote{It should be noted that this result
is \emph{not} true for the key itself.  If Eve only wants to learn if the
key was a particular value $k_0$, she may do so without disturbing
the state very much}.

Throughout this section we work with some group operator $+$ and
all operations are in that group.  In dimension $2^n$ the $+$ operator will
usually be bitwise exclusive or (XOR).

\begin{theorem}
\label{thm:ivd_fm}
Alice sends the $D$ dimensional state $\ket{k}$ to Bob, with $k$ chosen
uniformly at random, and after Bob has received the state Alice announces
$a=m + k$ (represented by the random variable A).
Denote $f(M)$ as the function $f$ of the random variable $M$,
and $f(K)$ is the function $f$ of the random variable $K$.
The
information Eve can get about any function of $m$, $f(m)$, is bounded
by the square root of the probability that Eve would have caused
errors in any MUB:
\begin{eqnarray*}
I(f(M);E|A)&\le& H(f(K))4\sqrt{P_{\widetilde{e}}}
\end{eqnarray*}
\end{theorem}
\prf
This proof will follow closely the proof of
theorem \ref{thm:ivd} and use the same tools.
If $a = m + k$, then $f(m)=f(a - k)$.
The state consistent with a function value $i$ is:
\begin{eqnarray*}
{\sigma_i}^a &\equiv& \frac{1}{q_i} \sum_{k:f(a - k)=i} p_k \rho_k
\end{eqnarray*}
with $q_i \equiv \sum_{k:f(a - k)=i} p_k$.
Note that since $p_k=\frac{1}{d}$, then the probability of an
announcement $a=m + k$ is also $\frac{1}{d}$.  As such, $q_i$ does not
depend on $m$ and is only related to the number of inputs to the function $f$
which have a given output.
The averaged state is:
\begin{eqnarray*}
\sigma^a &\equiv& \sum_i q_i {\sigma_i}^a\\
&=&\sum_i\sum_{k:f(a - k)=i}p_k\rho_k
\end{eqnarray*}
Since each input has one and only one output and $p_k = \frac{1}{d}$:
\begin{eqnarray*}
\sigma^a &=&\sum_k\frac{1}{d}\rho_k=\rho
\end{eqnarray*}
The definition of mutual information\cite{CoverThomas} means that:
\begin{eqnarray*}
I(f(M);E|A)&=&\sum_a p_a I(f(M);E|A=a)
\end{eqnarray*}
Using lemma \ref{lemm:nbitSD}
\begin{eqnarray*}
\lefteqn{\sum_a p_a I(f(M);E|A=a)}\\
&\le& -\sum_a p_a \sum_i q_i \log q_i |{\sigma_i}^a - {\sigma}^a|\\
&=& -\sum_i q_i \log q_i \sum_a p_a | {\sigma_i}^a - \rho|\\
&=& -\sum_i q_i \log q_i \sum_a p_a | {\sigma_i}^a -
\nproj{\widetilde{\phi_0}} + \nproj{\widetilde{\phi_0}} -
\rho|\\
&\le& -\sum_i q_i \log q_i \sum_a p_a
\left(| {\sigma_i}^a - \nproj{\widetilde{\phi_0}}|
+ |\nproj{\widetilde{\phi_0}} - \rho|\right)\\
&=& -\sum_i q_i \log q_i\sum_a p_a
\left(2 \sqrt{ 1 - \frac{ \bra{\widetilde{\phi_0}}{\sigma_i}^a\ket{\widetilde{\phi_0}} }
{ \braket{\widetilde{\phi_0}}{\widetilde{\phi_0}} }
}
+ 2 \sqrt{ 1 - 
\frac{ \bra{\widetilde{\phi_0}}\rho\ket{\widetilde{\phi_0}} }
{ \braket{\widetilde{\phi_0}}{\widetilde{\phi_0}} }
}\right)\\
&\le& -\sum_i q_i \log q_i
\left(2 \sqrt{ 1 - 
\frac{ \bra{\widetilde{\phi_0}}\sum_a p_a {\sigma_i}^a\ket{\widetilde{\phi_0}} }
{ \braket{\widetilde{\phi_0}}{\widetilde{\phi_0}} }
}
+ 2 \sqrt{ 1 - \frac{ \bra{\widetilde{\phi_0}}\rho\ket{\widetilde{\phi_0}} }
{ \braket{\widetilde{\phi_0}}{\widetilde{\phi_0}} }
}\right)
\end{eqnarray*}
We can simplify the quantity $\sum_a p_a {\sigma_i}^a$ by remembering that $p_a
= 1/d$ and $q_i$ is independent of $a$:
\begin{eqnarray*}
\sum_a \frac{1}{d}{\sigma_i}^a &=&\sum_a \frac{1}{d}
\frac{\sum_{k:f(a - k)=i}\frac{1}{d}\rho_k}{q_i}\\
&=&\frac{1}{q_i}\sum_a \frac{1}{d}\sum_{m:f(m)=i} \frac{1}{d}\rho_{a +
m}\\
&=&\frac{1}{q_i}\sum_{m:f(m)=i}\frac{1}{d}\sum_a \frac{1}{d}\rho_{a +
m}
\end{eqnarray*}
In the last sum, we sum over all $a$ with equal weight; hence, the $m$
dependence disappears:
\begin{eqnarray*}
\sum_a \frac{1}{d}{\sigma_i}^a &=& \frac{1}{q_i}\sum_{m:f(m)=i}\frac{1}{d}\sum_a \frac{1}{d}\rho_{a + m}\\
&=& \frac{1}{q_i}(\sum_{m:f(m)=i}\frac{1}{d}) \rho\\
&=& \rho
\end{eqnarray*}
Putting this back into the information bound:
\begin{eqnarray*}
\lefteqn{\sum_a p_a I(f(M);E|A=a)}\\
&\le& -\sum_i q_i \log q_i
\left(2 \sqrt{ 1 - 
\frac{ \bra{\widetilde{\phi_0}}\sum_a p_a {\sigma_i}^a\ket{\widetilde{\phi_0}} }
{ \braket{\widetilde{\phi_0}}{\widetilde{\phi_0}} }
}
+ 2 
\sqrt{ 1 - \frac{ \bra{\widetilde{\phi_0}}\rho\ket{\widetilde{\phi_0}} }
{ \braket{\widetilde{\phi_0}}{\widetilde{\phi_0}} }
}\right)\\
&=&-\sum_i q_i \log q_i
(4 \sqrt{ 1 - \frac{ \bra{\widetilde{\phi_0}}\rho\ket{\widetilde{\phi_0}} }
{ \braket{\widetilde{\phi_0}}{\widetilde{\phi_0}} }
})\\
&=& 4 H(Q) \sqrt{ 1 - \frac{ \bra{\widetilde{\phi_0}}\rho\ket{\widetilde{\phi_0}} }
{ \braket{\widetilde{\phi_0}}{\widetilde{\phi_0}} } }\\
&\le & H(f(K))4\sqrt{ P_{ \widetilde{ e } } }
\end{eqnarray*}
Which proves the result.
\qed

\section{Concluding Remarks}
By developing bounds on entropy, we are able to bound
the amount of
information that measurements can get from a quantum source.  Modeling
eavesdropping in quantum key distribution as a quantum source, we are able to
bound information that an eavesdropper can get.  Since this bound is a
function of the errors that would be caused in any MUB,
Alice and Bob can use their measurements to estimate this figure. 
Therefore, Alice and Bob can
bound information that Eve has about the information they share. 
In addition to showing security of such information, we show
that any function of messages encrypted with this secret information
is secure.  
This is a very strong statement about the robustness of quantum security.

\nonumsection{References}

\bibliographystyle{unsrt}
\bibliography{infodist}

\newpage 

\appendix

\section{Bound on Mutual Information for 1-bit Sources}
Suppose there is a classical source $S$ which sends one of two signals;
zero or one.
Also suppose that $p_{s=1}\le p_{s=0}$.
Following \cite{fuchs99}, we first come up with a linear bound on $H(p)$:
\begin{lemma}
\label{lemm:entropy-bound}
For any concave function $H(p)$ with $H(0)=H(1)=0$
and any $p'\le 1/2$, $H(p)\ge H(p') - \frac{H(p')}{p'}|p-p'|$
\end{lemma}
\prf
Consider two regions, $p \le p'$ and $p > p'$.  $H(p)$ is concave, which means
that $H(\alpha x + (1-\alpha)y) \ge \alpha H(x) + (1-\alpha)H(y)$.
Applying this with $x=p'$, $\alpha = p/{p'}$ and $y=0$, we obtain:
$H(p) \ge \frac{H(p')}{p'}p$, which is exactly what we need for $p \le p'$.
In the region $p > p'$ we want to show that $H(p) \ge H(p') - \frac{p-p'}{p'}H(p')$.
Again using the concavity, set $y=p',x=1$ and $\alpha = \frac{p - p'}{1-p'}$
We see then that
\begin{eqnarray*}
H(p) &=& H(\frac{p - p' + p'- pp'}{1-p'})\\
                         &=& H(\frac{p - p'}{1-p'} + \frac{1 - p}{1-p'}p')\\
			 &\ge&  \frac{p - p'}{1-p'} H(1) + \frac{1 - p}{1-p'}H(p')\\
			 &=& \frac{1 - p}{1-p'} H(p')\\
			 &=& H(p') - \frac{p - p'}{1-p'}H(p')
\end{eqnarray*}
Since $p' \le 1/2$, this implies that
$\frac{1}{1-p'} \le 2 \le \frac{1}{p'}$ and $\frac{-1}{1-p'} \ge \frac{-1}{p'}$.
We know that $p > p'$ in this region, so $p-p'$ is positive, thus:
\begin{eqnarray*}
H(p)&\ge& H(p') - \frac{p - p'}{1-p'}H(p')\\
    &\ge& H(p') - \frac{p - p'}{p'}H(p')
\end{eqnarray*}
\qed

\begin{lemma}
The mutual information between the random variable $E$ and the random
bit $S$ (with $p(s=0) \ge p(s=1)$) is bounded:
\begin{eqnarray*}
I(E;S)\le H(S)p(s=0)\sum_e|p(e|s=1) - p(e|s=0)|
\end{eqnarray*}
\end{lemma}
\prf
Using lemma \ref{lemm:entropy-bound} as a bound on $H(S|E)$ with
$p'=p(s=1)$, we can obtain the bound on mutual information:
\begin{eqnarray*}
I(E;S)&=&H(S)-H(S|E)\\
&=&H(S) - \sum_e p_{e}H(S|E=e)\\
&\le&H(S) - \sum_e p_{e} (H(p(s=1)) -
\frac{H(S)}{p(s=1)}|p(s=1|e) - p(s=1)|)\\
&=& H(S)\sum_e |p(e|s=1) - p(e)|\\
&=& H(S)\sum_e |p(e|s=1) - (p(s=0) p(e|s=0) +
p(s=1)p(e|s=1))|\\
&=& H(S)p(s=0)\sum_e | p(e|s=1) - p(e|s=0)|
\end{eqnarray*}
\qed

\begin{lemma}
\label{lemm:1bitSD}
If a source $S$ outputs quantum states $\rho_0$ and $\rho_1$ with
probabilities $p_0$ and $p_1$ with $p_0 \ge p_1$, then mutual information
between this source and the output of any measuring device $E$ is bounded:
$I(E;S)\le H(S)p(s=0) Tr|\rho_0 - \rho_1|$
\end{lemma}
\prf
The source sends two states, $\rho_0$ and $\rho_1$.  Eve does some
POVM\cite{peres} on them.  The probability that Eve gets outcome $x$ for her
measurement given an input $s$ is: $p(e|s)=Tr(E_e \rho_s)$.  This gives:
\begin{eqnarray*}
I(E;S)&\le& H(S)p(s=0)\sum_e | Tr(E_e(\rho_0 - \rho_1)) |
\end{eqnarray*}
Since $\rho_0 - \rho_1$ is Hermitian, we can diagonalize it as
$\sum_i \lambda_i \density{\psi_i}{\psi_i}$.  Taking this and applying
the facts that $E_e$ are positive semi-definite and $\sum_e E_e = I$, we get:
\begin{eqnarray*}
I(E;S)&\le& H(S)p(s=0)\sum_e | Tr(E_e(\rho_0 - \rho_1)) |\\
&=& H(S)p(s=0)\sum_e |Tr(E_e(\sum_i \lambda_i \density{\psi_i}{\psi_i}))|\\
&=& H(S)p(s=0)\sum_e | \sum_i \lambda_i \bra{\psi_i}E_e\ket{\psi_i}|\\
&\le& H(S)p(s=0) \sum_e \sum_i |\lambda_i| \bra{\psi_i}E_e\ket{\psi_i}\\
&=& H(S)p(s=0) \sum_i |\lambda_i| \bra{\psi_i}\sum_e E_e\ket{\psi_i}\\
&=& H(S)p(s=0) \sum_i |\lambda_i|\\
&=& H(S)p(s=0) Tr|\rho_0 - \rho_1|
\end{eqnarray*}
\qed

\begin{corollary}
If a source $S$ outputs quantum states $\rho_0$ and $\rho_1$,
then mutual information between this source and the output of any measuring
device $E$ is bounded:
$I(E;S)\le H(S) Tr|\rho_0 - \rho_1|$
\end{corollary}
\prf
Consider two cases, the first where $p_0 \ge p_1$ and
the second where $p_1 > p_0$.
If $p_0 \ge p_1$, then using lemma \ref{lemm:1bitSD} we have that
$I(E;S)\le H(S)p(s=0) Tr|\rho_0 - \rho_1|$.  Since $p(s=0)\le 1$,
we get the result.  If $p_1 > p_0$ then relabel the $\rho_1$ as $\rho_0$
and vice versa.  Hence in the original labeling, lemma \ref{lemm:1bitSD}
becomes
\begin{eqnarray*}
I(E;S)\le H(S)p(s=1) Tr|\rho_1 - \rho_0|
\end{eqnarray*}
, and since $p(s=0)\le 1$
we get the result.
\qed

\section{Bounding the Trace Norm}
As we have seen in the previous section,
 the trace norm distance between quantum states is a powerful tool
for bounding mutual
information.  Now we look at some bounds on trace norm distances. 
\begin{lemma}
The trace norm distance between two pure states is:
\begin{eqnarray*}
|\density{\psi}{\psi} - \density{\phi}{\phi}| = 2\sqrt{1 -
|\braket{\psi}{\phi}|^2}
\end{eqnarray*}
\end{lemma}
\prf
Define $\braket{\psi}{\phi}=\alpha$.
Defining a new orthonormal basis we can write:
\begin{eqnarray*}
\ket{e_0}&\equiv&\ket{\psi}\\
\ket{e_1}&\equiv&\frac{1}{\sqrt{1-|\alpha|^2}}(\ket{\phi}-\alpha\ket{\psi})
\end{eqnarray*}
Inverting these equations we have:
\begin{eqnarray*}
\ket{\psi} &=& \ket{e_0}\\
\ket{\phi} &=& \alpha\ket{e_0}+ \sqrt{1-|\alpha|^2}\ket{e_1}
\end{eqnarray*}
Using this new basis, we find that:
\begin{eqnarray*}
|\density{\psi}{\psi} - \density{\phi}{\phi}|
&=& |(1-|\alpha|^2)\density{e_0}{e_0}
- (1-|\alpha|^2)\density{e_1}{e_1}\\
&&- \sqrt{1-|\alpha|^2}(\alpha^{*}\density{e_1}{e_0}+ \alpha\density{e_0}{e_1})|
\end{eqnarray*}
This is just a $2\times 2$ matrix and we can compute the trace norm
by taking the absolute value of the eigenvalues, which are:
\begin{eqnarray*}
\lambda&=&\stackrel{+}{-}\sqrt{1 - |\alpha|^2}
\end{eqnarray*}
\qed

\begin{lemma}
\label{lemm:trace_bound_mixed_pure}
The trace norm distance between any state and any pure state is bounded:
\begin{eqnarray*}
|\rho - \density{\psi}{\psi}| \le 2\sqrt{1 - \bra{\psi}\rho\ket{\psi}}
\end{eqnarray*}
\end{lemma}
\prf
Let $\rho = \sum_i p_i \density{\phi_i}{\phi_i}$ and apply $\sum_i p_i x_i \le
\sqrt{\sum_i p_i {x_i}^2}$:
\begin{eqnarray*}
|\rho - \density{\psi}{\psi}| &=&
|\sum_i p_i \density{\phi_i}{\phi_i} - \density{\psi}{\psi}|\\
&\le& \sum_i p_i |\density{\phi_i}{\phi_i} - \density{\psi}{\psi}|\\
&=& \sum_i p_i \sqrt{1 -|\braket{\psi}{\phi_i}|^2}\\
&\le& \sqrt{ \sum_i p_i (1 -|\braket{\psi}{\phi_i}|^2)}\\
&=& 2\sqrt{1 - \bra{\psi}\rho\ket{\psi}}
\end{eqnarray*}
\qed

\begin{definition}
\label{def:purification}
Purification of $\rho$: any pure state $\ket{\psi}$ in 
$\mathcal{H}_{1}\otimes\mathcal{H}_{2}$
such that $Tr_2(\density{\psi}{\psi})=\rho$
\end{definition}

\begin{lemma}
\label{lemm:trace_out_bound}
The trace norm distance is reduced by partial trace:
\begin{eqnarray*}
|\rho' - \sigma'| &\le& |\rho - \sigma|
\end{eqnarray*}
Where $\rho$ and $\sigma$ are density matrices over states in ${\mathcal H}_1
\otimes {\mathcal H}_2$ and the partial trace is over one of the subsystems:
$\rho' = Tr_2(\rho)$ and $\sigma'=Tr_2(\sigma)$.
\end{lemma}
\prf
See \cite{peres}.
\qed

\end{document}

%% file: basic_attack.eepic
\setlength{\unitlength}{0.00087489in}
\begingroup\makeatletter\ifx\SetFigFont\undefined%
\gdef\SetFigFont#1#2#3#4#5{%
  \reset@font\fontsize{#1}{#2pt}%
  \fontfamily{#3}\fontseries{#4}\fontshape{#5}%
  \selectfont}%
\fi\endgroup%
{\renewcommand{\dashlinestretch}{30}
\begin{picture}(3180,1839)(0,-10)
\path(1365,1812)(2265,1812)(2265,12)
	(1365,12)(1365,1812)
\path(465,1362)(1365,1362)
\blacken\path(1245.000,1332.000)(1365.000,1362.000)(1245.000,1392.000)(1245.000,1332.000)
\path(465,462)(1365,462)
\blacken\path(1245.000,432.000)(1365.000,462.000)(1245.000,492.000)(1245.000,432.000)
\path(2265,1362)(3165,1362)
\blacken\path(3045.000,1332.000)(3165.000,1362.000)(3045.000,1392.000)(3045.000,1332.000)
\path(2265,462)(3165,462)
\blacken\path(3045.000,432.000)(3165.000,462.000)(3045.000,492.000)(3045.000,432.000)
\put(15,1362){\makebox(0,0)[lb]{\smash{{\SetFigFont{12}{14.4}{\rmdefault}{\mddefault}{\updefault}$\ket{i}$}}}}
\put(15,462){\makebox(0,0)[lb]{\smash{{\SetFigFont{12}{14.4}{\rmdefault}{\mddefault}{\updefault}$\ket{0}$}}}}
\put(1590,912){\makebox(0,0)[lb]{\smash{{\SetFigFont{12}{14.4}{\rmdefault}{\mddefault}{\updefault}$U_{Eve}$}}}}
\put(3165,912){\makebox(0,0)[lb]{\smash{{\SetFigFont{12}{14.4}{\rmdefault}{\mddefault}{\updefault}$\sum_j\ket{E_{i,j}}\ket{j}$}}}}
\end{picture}
}